# Time-efficient Garbage Collection in SSDs

Lars Nagel
*Loughborough University*

Tim Süß, Kevin Kremer, M. Umar Hameed, Lingfang Zeng, André Brinkmann
*Johannes Gutenberg University Mainz, Germany*

## Abstract

SSDs are currently replacing magnetic disks in many application areas. A challenge of the underlying flash technology is that data cannot be updated in-place. A block consisting of many pages must be completely erased before a single page can be rewritten. This victim block can still contain valid pages which need to be copied to other blocks before erasure. The objective of garbage collection strategies is to minimize write amplification induced by copying valid pages from victim blocks while minimizing the performance overhead of the victim selection.

Victim selection strategies minimizing write amplification, like the cost-benefit approach, have linear runtime, while the write amplifications of time-efficient strategies, like the greedy strategy, significantly reduce the lifetime of SSDs. In this paper, we propose two strategies which optimize the performance of cost-benefit, while (almost) preserving its write amplification. Trace-driven simulations for single- and multi-channel SSDs show that the optimizations help to keep the write amplification low while improving the runtime by up to 24-times compared to the original cost-benefit strategy, so that the new strategies can be used in multi-TByte SSDs.

## 1 Introduction

SSDs are faster and more reliable than hard disks which mechanically move their read/write heads, but the cost per storage capacity is still higher for SSDs. The reliability and performance advantages of SSDs have nevertheless led to a wide-spread adoption of SSDs in tablets, notebooks, servers, and enterprise storage [5].

The shift from hard disks to flash-based solid state drives (SSDs) is a huge change in the architecture of storage devices. The underlying NAND flash introduces, besides all of its advantages, the challenge that data cannot be overwritten in place (with the exception of using WOM codes [17]). The flash translation layer (FTL), managing the mapping of virtual pages to physical pages, has to erase a complete victim block before a single page in this block can be overwritten. For this reason, the garbage collection (GC) first copies every valid page of a victim block to a clean block. These out-of-place updates of pages are very similar to the management of log-structured file systems (LFS) in which data are written to a new location instead of being written in place.

Several strategies for choosing victim blocks have been proposed with the following three objectives:

- Small write amplification (WA): The number of clean pages of victim blocks should be small (on average) to minimize the *write amplification*.

- Low search cost: The *complexity* to search for the next victim block should not grow linearly (or worse) in the number of blocks to keep the selection time small.

- Balanced hardware wear-out: The *wear-leveling* ensures that each block is erased approximately the same number of times.

Many GC strategies perform well for uniform access distributions where the probability of a write access hitting a page is the same for all pages. Interestingly, it can be proven that the greedy algorithm, which simply selects the blocks with the minimum number of valid pages, performs best in this setting [8]. Other simple cleaning policies like least recently written (LRW) also perform well in this case.

In the case of non-uniform page access distributions, some blocks are more frequently accessed. These blocks are called hot blocks and less frequently accessed blocks are cold blocks. The difference of the access frequencies disturbs the performance of simple cleaning policies and write amplification and wear-leveling get compromised [23].

A prominent strategy for real-world distributions is the *cost-benefit approach* [12]. Cost-benefit (CB) includes two parameters, the time since the last page invalidation and the number of valid pages. The strategy ensures that cold blocks are selected sooner, effectively leading to better overprovisioning. The disadvantage of this strategy is its selection time due to the need of computing the CB value for all blocks at every GC-cycle, which leads to a time complexity of $O(n)$ to select one out of $n$ blocks.

Progress on flash technology raised SSD capacities up to 60 TByte. While increasing the capacity, there are the possibilities to add more pages to a block or to increase the number of blocks. In terms of GC efficiency, both cases induce tradeoffs between WA and search cost for victim blocks, making linear time algorithms infeasible. In this paper we introduce several optimizations of the original cost-benefit approach to improve its performance and to allow its usage for larger SSDs.

- The *fast cost-benefit* strategy orders blocks in sets called *classes*. Few blocks with high CB values are kept in the highest priority set, where victim blocks are selected from. Blocks from lower-value classes are moved to higher-priority classes if their CB value passes a threshold. The time when a block has to be moved can be calculated in advance.

- The *approximative cost-benefit* approach does not individually select victim blocks, but always calculates a bunch of possible future victim blocks, reducing the calculation time by a constant factor.

The approaches are not restricted to improve the CB algorithm, but can also be generalized to many additional victim selection strategies like FeGC or CAT [14][4].

We evaluate the improvements for different SSD sizes, access patterns, and numbers of pages per block by comparing their performance and write amplification with well-known strategies: greedy, CAT, FeGC and the original cost-benefit approach. We show that the new strategies achieve a write amplification at least as good as cost-benefit, CAT, and FeGC, while being significantly faster, both for Intel server processors as well as for embedded ARM processors. The strategies therefore allow vendors to build SSDs applying sophisticated GC strategies improving SSD lifetimes for multi-TByte capacities at a performance far beyond one million IOPS.

The rest of the paper is organized as follows. We start with a discussion related work in Section 2. In Section 3, we present our new algorithms and analyze their expected runtimes. Using a trace-driven simulation tool, we then evaluate old and new strategies in Section 4 before we conclude the paper in Section 5.

## 2  Related Work

For technical reasons, SSDs do not erase single pages, but whole blocks which contain a large number of pages. When there is no space left for writing new pages, the garbage collection process residing in the FTL has to determine a block or blocks for cleaning.

A block's life-cycle is depicted in Fig. 1. Physical pages are written to a clean, active block in a log-structured manner. Data is never erased from a fully written block before it is selected for GC again. Instead pages that are no longer valid are marked as *invalid*. At first, a block has $N_p$ valid pages assuming that $N_p$ is the number of pages per block and that no page has been marked invalid while the block was written. Over time, the number of valid pages decreases. When the block is selected for cleaning, its remaining valid pages are copied to the now active blocks before the data are erased (see, e.g. [15]).

In many scenarios, it makes sense to distinguish between hot and cold data and to have one active block (*i.e.*, a block that is currently written) for each predefined "temperature" range [9][6][21][10]. Then a new page of data that, due to its past behavior, is known to be cold or hot can be placed on the respective active block.

Efficient block selection for cleaning flash memory has a substantial impact on the storage layer, and the applied GC strategies strongly depend on the mode to translate virtual addresses to page addresses. In this paper we focus on page mapping strategies, where each page address can be mapped to each block. Therefore the spacial locality of pages is not preserved within the blocks. Related pages are scattered among all blocks. This distribution however allows a better utilization of the SSD's capacity. This address translation requires more DRAM memory than block mapping or hybrid mapping schemes [13][7][16], but can also deliver substantially better write amplification results.

However, up to now few strategies have worked on improving the performance of selecting victim blocks while achieving very good WA and wear-leveling.

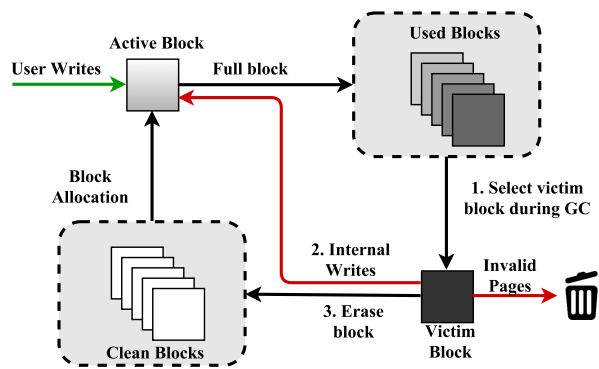

Figure 1: Life-cycle of a block (adapted from [15])

A simple approach is the **greedy algorithm**, which has been proposed, *e.g.*, by Wu et al. [23]. The greedy strategy selects the block with the least number of valid pages as victim block. It keeps (in a simple implementation) a heap containing all full blocks which is updated in $O(\log(n))$ at every page invalidation. The victim selection can also be implemented in constant time keeping an array of lists with one list for every possible number of valid pages in a block. Greedy achieves an optimal WA for uniform access patterns [8]. However, for other access patterns it is far from being optimal, and it does not take wear-leveling into account.

The **cost-benefit approach** (CB) for victim selection was proposed by Rosenblum *et al.* [19] in the context of log-structured file systems and has been adapted by Kawaguchi *et al.* [12] for flash storage. CB considers the block's utilization and its age. It picks the block $b$ with the highest cost-benefit value

$$f_b^{cb}(t) = a_b(t) \cdot \frac{1 - u_b(t)}{2 \cdot u_b(t)} \quad (1)$$

at the cleaning time $t$, where the age $a_b(t)$ is the time since the last page modification and the utilization $u_b(t)$ is the fraction of valid pages in $b$. The term $2 \cdot u_b(t)$ represents the cost as $u_b(t) \cdot N_p$ valid blocks have to be read and written during cleaning. The benefit is proportional to $(1 - u_b(t))$ because $(1 - u_b(t)) \cdot N_p$ pages are cleaned during GC. The age $a_b(t)$ helps to GC cold blocks as their pages are typically invalidated very slowly.

CB has a good write amplification and wear-leveling, but its victim selection is slow because $f_b^{cb}(t)$ has to be computed for all $n$ blocks, and due to its time dependence the computation can only be started when GC is invoked. Our strategy described in Section 3 improves CB performance while preserving WA and wear-leveling.

The **cost-age-times strategy** (CAT) by Chiang *et al.* [4] follows an approach similar to CB, but also includes the number of times a block has been erased. It determines the block with the highest function value

$$f_b^{cat}(t) = \frac{1 - u_b(t)}{u_b(t)} \cdot age_b \cdot \frac{1}{erase\_count}. \quad (2)$$

Here the $age_b$ is defined as the time elapsed since block creation. The *age* of a block is further normalized by a transformation function only including several discrete values, which nearly resembles a log-function. (In the CB term the constant factor $1/2$ is left out because it has no effect when comparing blocks.)

Menon *et al.* [18] proposed an **age-threshold algorithm** for garbage collection in *log-structured file systems* (LFS) that also considers age and utilization, but in a simpler way. It preselects all blocks that are older than a certain age and then picks a block with the least number of valid pages from them. A drawback of this approximation is that the preselection does not consider utilization of blocks. Finding a suitable age threshold is furthermore critical, as no threshold fits all cases.

Kwon *et al.* [14] suggested the **fast and endurant garbage collection** (FeGC) which selects the victim block based on its complete invalidation history. FeGC computes the *cost with age* (CwA) value of each block by summing up the "ages" of all its $n_{inv}$ invalid pages,

$$f_b^{cwa}(t) = \sum_{i=1}^{n_b^{inv}} (t - t_i), \quad (3)$$

where the age of page $i$ is the difference between the current time $t$ and the time $t_i$ at which it became invalid. FeGC selects the block with the largest CwA value. The strategy prefers blocks with many invalid pages, but also regards cold pages with *old* invalid pages.

Additional GC strategies are described, e.g., in [11], [24], and [3]. These strategies include wear-leveling strategies, are optimized for specific file systems or especially support multi-channel SSDs.

**Wear Leveling Approaches**

There are numerous algorithms that focus on uniform wear-out of blocks to equalize P/E cycles. A simple approach is selecting a block in FIFO manner, i.e., the first block written is selected first for GC. FIFO does not achieve a good WA because in many cases the oldest block can still have a large number of valid pages.

An even wear-leveling of blocks can be, e.g., achieved by the dual-pool (DP) algorithm of Chang *et al.* [2]. DP partitions blocks into two pools, a hot and cold pool. For wear-leveling, the algorithm regularly checks the erase count difference between the hottest block with the highest erase count with the coldest block with the lowest erase count and exchanges them in case that this difference is above a predefined value. DP does not include measures to minimize the write amplification.

Ban proposed a solution to improve wear-leveling in flash that ensures that static areas of cold blocks can be moved to other locations. The wear-leveling algorithm can either be started when a large number of writes has occurred or according to a random process. The algorithms then just selects a block (e.g., by just using a counter), which can be either hot or cold. If the process detects a cold block (called static one) then it exchanges this with a hot region, so that the block can take part in the wear-leveling. The advantage of the process is that every block will at least take part in this process a few hundred times over the lifetime of an SSD and therefore an even balancing of erase operations can be ensured [1].

The strategies proposed in [2] and [1] show that wear-leveling can efficiently be achieved orthogonal to GC.

We therefore do not further consider wear-leveling and assume that the GC strategies can be coupled with a good wear-leveling strategy.

## 3 Victim Selection Methods

In this section we describe two modified cost-benefit strategies to select a victim block. The first strategy, called *fast cost-benefit*, determines the victim like the original cost-benefit strategy, but accelerates the process by grouping blocks in classes and, by this means, reduces the search time. Blocks with high cost-benefit values are assigned to Class 0 from which the victim is chosen, and the remaining blocks are stored in Class 1. Blocks are moved between classes if their value passes a threshold.

The *approximative cost-benefit* strategy regularly builds a set of $q$ candidate blocks from which the next $q$ victim blocks are drawn. When the set is empty, a new set is built. In contrast to the fast cost-benefit strategy, a block is not moved in or out of the set when its CB value changes. Hence, it is not ensured that approximative cost-benefit always selects the block with the best CB value. The advantage of the strategy is that the linear CB search time can be significantly reduced by running it $q$-times less frequent than in the original algorithms. The drawback is that it might worsen the write amplification.

### 3.1 Fast Cost-benefit Strategy

The aim of the fast cost-benefit strategy is to reduce the victim selection time. The blocks are therefore kept in two disjoint classes, Class 1 and Class 0. The classes contain only fully written blocks and each page in a block is either (written and valid) or (written and invalid). Class 1 contains all blocks whose CB value is smaller than a threshold $t_{cb}$, which can be automatically adjusted at runtime. Class 0 contains all blocks whose CB value is greater than $t_{cb}$.

With this class structure, the victim selection becomes more efficient because all candidates are in Class 0 (unless Class 0 is empty) and blocks with smaller CB values need not be considered. If the number of blocks in Class 0 can be kept small or even constant, then the victim block search can be much faster than the linear search of the standard CB strategy.

The classification of the blocks comes at the obvious price that the classes have to be kept up-to-date. A shift from Class 1 to Class 0 is triggered when the value of the (original) cost-benefit function (see Equation 1) becomes larger than the threshold $t_{cb}$ due to the age $a_b(t)$ of a block $b$. Block aging requires a data structure called `shifts` which stores for each block in Class 1 at which time it has to be moved to Class 0, provided that the block's utilization does not change. In our implementation, we keep the respective timestamps sorted and move the blocks with timestamps smaller than the current time to Class 0 whenever a victim is to be selected.

Unfortunately, blocks do not always move forward, *i.e.*, from Class 1 to Class 0, but also the other way round. When the utilization of a block $b$ is changed, its age is reset to 0 because the age $a_b(t)$ is defined as the difference between the current time and the time of the last invalidation. The consequence is that the function value also becomes 0 and that the block is moved to Class 1. The required shifts or updates can be easily handled in the function invalidating the page, but lead to additional bookkeeping overhead.

Figure 2 shows an example of $t_{cb} = 15,000$ and 256 pages per block. Class 1 on the left contains all blocks with a CB value smaller than 15,000. Block $\mathbf{b_7}$ with its age of 24,469 has reached a cost-benefit value of 15,001 and must therefore be moved to Class 0. For the reverse case in which a block moves from Class 0 to Class 1 consider, for instance, the invalidation of a page for $\mathbf{b_{55}}$. Its age, *i.e.* the difference between the current time and the time of the last invalidation, would be set to 0 and $\mathbf{b_{55}}$ would be moved back to Class 1. Finally, when the GC process searches Class 0, it picks block $\mathbf{b_8}$ because it has the largest cost-benefit value.

Compared to the normal cost-benefit strategy, we make the following changes to the code:

- A block starts in Class 1 after its last page has been written and the block is moved to the set of used blocks (see Figure 1).

- The age of a block is always set to 0 when a page is

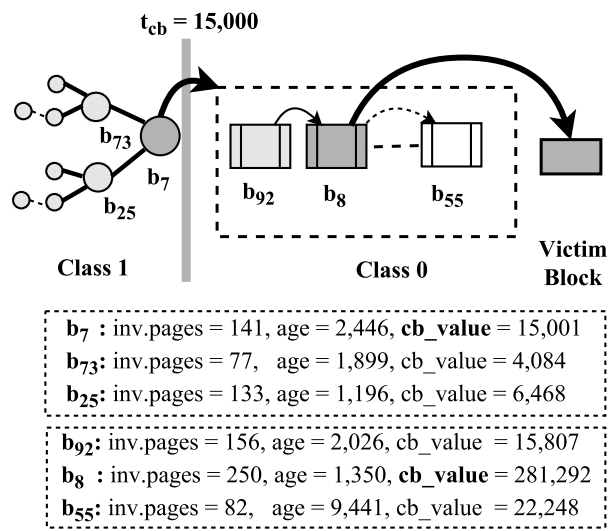

Figure 2: Class structure of the fast cost-benefit strategy. The threshold $t_{cb}$ is set to $t_{cb} = 15,000$.

invalidated. Hence, if the block is in Class 0, it has to be moved to Class 1. Independent of its previous class, it is necessary to update its shift time based on the new utilization and to find its position in Class 1.

The function that determines the victim block is considerably changed and extended; its pseudo code is given in Algorithm 1.

- Before a victim is selected for GC, the blocks in Class 1 which are past their shifting time, i.e. whose CB value is greater than the threshold $t_{cb}$, are moved to Class 0 (lines 3-9).

- After that, it is checked if the number of blocks in Class 0 is greater than 0 and smaller than $t_0$ (here set to 125). If it is empty or has more than $t_0$ blocks, the threshold is adapted and blocks are shifted such that Class 0 contains exactly $c_0$ blocks (lines 11-14), here set to 25.

These adaptations are necessary to keep Class 0 populated as well as small. The values for the maximum size $t_0$ of Class 0 and its initial size $c_0$ were determined in experiments and are a compromise between avoiding adaptions and facilitating quick searches. For efficiency, the adaptations can be merged with the subsequent victim selection, as done in our implementation.

- The actual victim selection is simple. For all blocks in Class 0, the CB value is computed and the block with the largest value is returned (lines 16-19).

The data structure of Class 0 is simply a list of blocks. Class 1 is represented by the data structure `shifts` (called *class1* in Alg. 1) for which we use C++'s `std::map`, basically a red-black tree providing sorting (needed for the lazy update of blocks) as well as quick inserts and deletes. For an implementation on a real SSD the STL containers can be exchanged by more space efficient data structures. As we want to use every timestamp only once, we increment the timestamp of a block in case of a collision. To create space for that, we first multiply each timestamp with a TIME_FACTOR.

## 3.2 Complexity of CB and Fast CB

In order to choose a victim, the standard CB strategy has to go through all blocks, compute their CB value and return the one with the largest value. So, if $n$ is the number of blocks, the time needed per victim is $O(n)$.

For our fast cost-benefit approach, we can estimate the time for victim selection in the same way. Let $C_i(t)$ denote the size of Class $i$ at time $t$ and let $C_i^{max} = \max_t\{C_i(t)\}$. Then the search for the victim alone takes $O(C_0(t))$ at time $t$.

**Algorithm 1** Function **get_victim_block()**

1: int ct, shift_time, cb, max_cb := 0
2: Block block := **null**
3: /* *Lazy update of blocks (cl. 1 → cl. 0)* */
4: ct = TIME_FACTOR * current_time
5: **while** class1 is not empty **do**
6:    (shift_time, block) := class1.next()
7:    **if** shift_time > ct **then**
8:       **break**
9:    move block from class1 to class0
10:
11: /* *Adjust threshold if class0 too large or small* */
12: **if** |class0| = 0 **or** |class0| > 125 **then**
13:    determine new threshold such that |class0| = 25
14:    reorganize blocks accordingly
15:
16: /* *Determine victim block* */
17: block := find_block_with_highest_cb_value(class0)
18: remove block from class0
19: **return** block

However, there is additional effort needed to keep up the class structure. Let $N_p$ be the number of pages per block and assume that the threshold is never adjusted. The shifting time needs to be adjusted at every page removal, compared to the threshold $t_{cb}$ and possibly moved to the other class. In total, these costs can be upper-bounded by $O(N_p \cdot \log C_1^{max})$ over the block's lifetime. The log-term is the time for updating one entry in the red-black tree and $N_p$ is the maximum number of times a page can be invalidated in a block. So, the lifetime cost for one block selected at time $t$ can be upper-bounded by:

$$O\left(C_0(t) + N_p \cdot \log C_1^{max}\right).$$

Since $C_1^{max} \leq n$ and $C_0(t)$ is bound by a constant, we can simplify the equation to

$$O(N_p \cdot \log(n)).$$

If $N_p \ll n/\log(n)$, our approach is asymptotically faster than the normal cost-benefit approach.

So far we ignored the adjustment of the threshold $t_{cb}$. If the adjustment is caused by Class 0 containing more than $t_0 = 125$ blocks, we must move $C_0(t) - c_0$ blocks to Class 1 and add them to `shifts`. Finding the $c_0 = 25$ blocks with the largest CB values is done by adding the first 25 blocks to a priority queue (or list), going through the remaining $C_0(t) - c_0 \approx 100$ blocks and replacing the block with the smallest cost-benefit value of the 25 blocks whenever a block with a larger CB value is found. The 25 blocks with the highest cost-benefit value then form the new Class 0, while the remaining $\approx 100$ blocks are moved to class Class 1, leading to additional costs of $O((t_0 - c_0) \log C_1(t))$. Because of $C_1(t) = \Theta(n)$, we can simplify this term to $O(\log n)$.

If the adjustment is caused by Class 0 being empty, we have to search Class 1 in order to move $c_0 = 25$ blocks to Class 0. This takes $O(n \cdot \log c_0) = O(n)$ operations.

Theoretically, these cases could happen often. Our experiments show that they are not too frequent, but also not rare. Conducting $9,893,523$ victim selections (using the TPCC trace described in Section 4.1), the number of adjustments was $51,050$. On average, in roughly 1 out of 194 victim selections Class 0 ran empty, and in roughly 1 out of 202 selections Class 0 had more than $t_0$ elements.

## 3.3 Approximative Cost-benefit Strategy

The approximative cost-benefit approach slightly alters the standard CB strategy to speed up victim selection – which might come at the cost of an increased write amplification. The approximative cost-benefit strategy does not individually select victim blocks, but creates a cache of $q$ blocks for future garbage collections.

When GC is invoked and the cache contains blocks, then the block with the highest CB value is chosen and evicted from the cache. Only in case of an empty cache, all blocks are scanned and the $q$ blocks with the highest CB value are moved to the cache.

The advantage is that one does not have to compute the CB values of all blocks for each garbage collection. Instead, $q - 1$ of $q$ GCs solely consider blocks from the cache, and only every $q^{th}$ run has to scan all blocks.

The disadvantage of the approach is that it does not always pick the highest-value block because, once the cache is filled, blocks outside of the cache are ignored even if their CB value has become better than the CB values of the blocks in the cache. This difference to the normal cost-benefit strategy can compromise the write amplification and is investigated in the next section.

## 4 Evaluation

In this section we present the evaluation of several victim selection strategies and compare the performance of our new approaches with the performance of existing strategies. For the evaluation, we use a trace-driven simulator written in C++ and three different real workload traces. The focus of the evaluation is on write amplification and the runtime of the GC algorithm. The evaluation does not consider the timings of flash chips or interfaces.

## 4.1 Simulator

SSDs consists of three major parts (see Fig. 3): an *Interface*, a *Controlling Unit* and a *Storage Unit*. The simulation environment focuses on the *Flash Translation Layer* (FTL) of the *Controller* which takes read and write requests from the *Interface*. The FTL maps logical to physical addresses and finds or creates empty pages when write requests arrive. The placement is based on the number of channels and write requests are distributed in a round-robin manner among them. The FTL is responsible to perform the GC as soon as the blocks attached to one channel run out of usable pages.

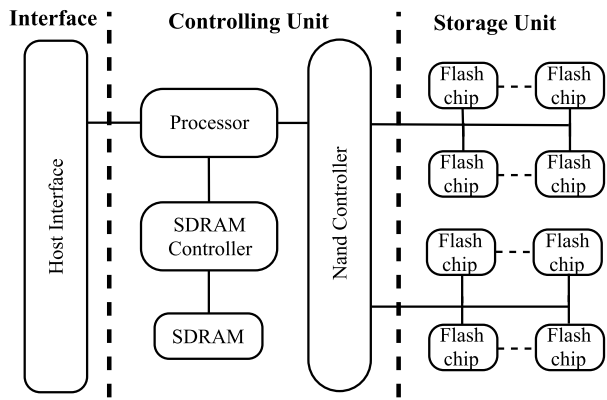

Figure 3: SSD Components

The Controller uses SDRAM as memory and accesses flash chips via the *NAND Controller*. We only investigate page mapping strategies where each page can be mapped to each block.

The simulator includes timings for translating logical to physical addresses, for selecting new victim blocks within GC, and for updating all internal data structures like heaps or lists. All dependent operations are mapped to a single CPU core; parallelization techniques have not been investigated. However, some algorithms can be parallelized, e.g., approximative cost-benefit allows to build a new set of $q$ potential victims before the current set is empty. Hence, the performance results form a **lower bound on the number of IOPS** which can be achieved by the evaluated algorithms. Technology-dependent timings for accessing pages or erasing blocks have not been added, but in many cases these operations can be carried out by an independent process on another CPU core. The write amplification results are accurately modeled. To analyze the impact of multiple channels on the tested strategies, we configured the simulated SSDs with either one or four channels. The simulator operates sequentially even when multiple channels are used.

The first trace is based on the *TPCC benchmark* which simulates a database environment and compares *online transaction processing* (OLTP) settings[1]. The traces were collected on an SSD with an address range of 500 GByte, a physical capacity of 600 GByte (including 20% overprovisioning) and a page size of 8 KByte. 11.323

---
[1] see http://www.tpc.org/tpcc/ for a description of the benchmark

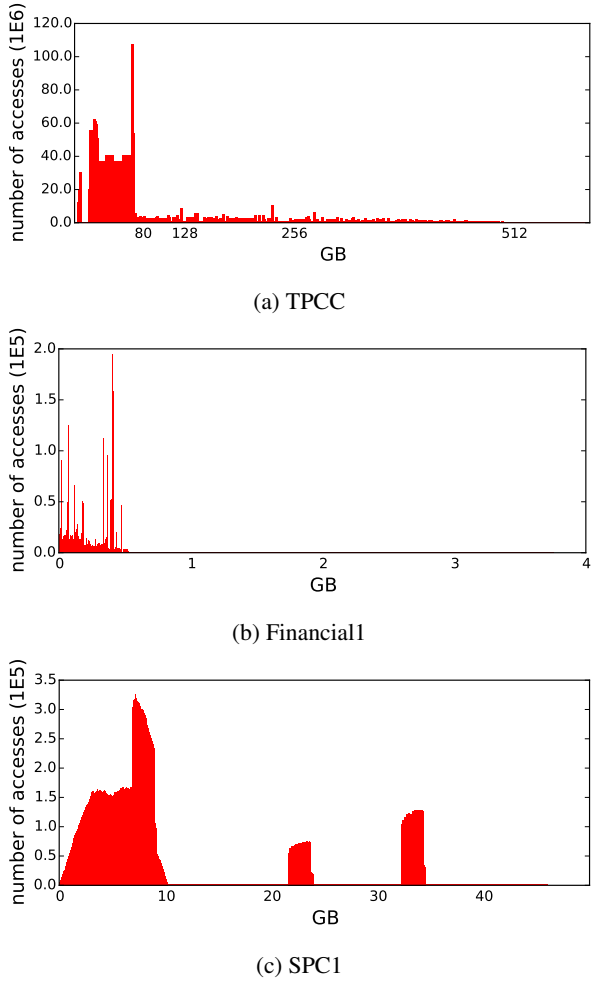

Figure 4: Access distribution of the applied traces

Tbyte of data were written while running the benchmark. The read accesses were removed from the traces because they have no influence on GC and WA. The access distribution is highly skewed; most accesses correspond to addresses below the 80 GByte boundary (see Figure 4a).

The simulations were also performed with the *Financial1* benchmark[2] and the *SPC-1*[3] trace. The general conclusions derived from these two benchmarks do not deviate from the ones of the TPCC benchmark so that we only partially present the respective simulations.

The Financial1 trace includes around 40,000,000 write requests accessing a 4 GByte SSD. The total amount of written data is about 310.6 GByte with an average write request size of 7.8 KByte. Almost all writes access the first 500 MByte of the SSD (see Fig. 4b).

The Storage Performance Council-1 (SPC-1) benchmark accesses an 80 GByte SSD. The trace contains 689,429,092 writes accessing 7.7 TByte in total. The trace has a variable requests sizes with an average of 11.2 Kbyte. Most writes are performed in one of three pools on the SSD in the first half of the SSD (see Fig. 4c).

GC is often regarded as independent from wear-leveling and hot/cold data classification. Nevertheless, achieving a good WA for heterogeneous access patterns requires a hot/cold data classification and, unless otherwise noted, we pre-characterized each (logical) page and assigned it to one of three hotness levels before running the simulation and we therefore provided three active blocks per channel. The characterization was based on the number of accesses to each page. The thresholds between the levels were manually selected[4]. The GC treated all blocks the same regardless of their hotness.

## 4.2 Strategies and Settings

We used the following settings for all simulations: The SSDs had capacities between 80 GByte and 1 TByte and the complete traces were processed, but accesses beyond the capacity of the simulated SSD were not performed. It is therefore possible that the number of accesses differed for different capacities. We applied an overprovisioning factor of 1.07, so that the physical capacity was 7% larger than the logical. The setting is close to the overprovisioning of consumer SSDs, while enterprise SSDs have bigger overprovisionings to improve WA and lifetime. Before each simulation, the SSD was warmed-up by sequentially writing the address space once, followed by random write accesses covering twice the capacity.

We used the fast cost-benefit strategy with the following parameters: The maximum number of blocks in Class 0 was $t_0 = 125$ and the number of blocks after re-adjustment $c_0 = 25$. The threshold $t_{cb}$ between Class 1 and 0 was adjustable and changed during the process.

The only parameter for the approximative cost-benefit strategy is $q$, the (maximum) number of victim blocks cached. We investigated the impact of $q$ on performance and write amplification.

The strategies taken from literature are greedy (using two data structures, one running in constant time, the other in logarithmic time), CB, CAT and FeGC. We show that CB is a good approach concerning WA and that its performance can be significantly improved by applying our optimizations. Strategies like CAT and FeGC suffer from the same (or even worse) performance overheads as CB, while not leading to smaller WAs. Greedy is an example of a strategy with constant, respectively logarithmic runtime, but bad WA for heterogeneous access patterns.

---

[2]see http://iotta.snia.org
[3]www.storageperformance.org

[4]The simulation environment and all parameters will be made available on GitHub.

The simulations were run on compute servers equipped with 64 Gbyte memory and two *Intel 2630v4* CPUs having 10 cores running at 2.2 GHz. All simulation used only a single core while no other programs were running on the same node. The operating system was *CentOS 7.3*, and the simulator was compiled with *GCC 5.4*. The workload traces were read from a Lustre 2.8 file system achieving 6 GByte/s bandwidth per node.

We also performed tests with an ARM Cortex-A17 Quad-Core SOC providing 1.8 GHz and 4 GByte RAM. The Financial1 trace was read from the internal eMMC flash memory which achieved a read bandwidth of 180 MByte/s. We have chosen this platform to investigate different CPU architectures and because SSDs are often equipped with similar embedded processors.

### 4.3 Write Amplification Results

The simulations recorded the the write amplification over time of the GC strategies for different SSD capacities, numbers of blocks, and page sizes.

Fig. 5 shows the WA development for the main algorithms (here still without approximative cost-benefit) over time for an SSD with a capacity of 1,024 GByte, 2,048 pages per block and a page size of 4 kByte. The simulation time on the x-axis depicts the real time when collecting the benchmark traces, as all accesses have been taken from the corresponding trace files. The start time at approx. 80,000 seconds is based on the warm-up phase, which has been simulated before the actual trace. No strategy has been able to adapt to the page hotness values until then and all GC strategies need some time to adapt to the new access patterns of the TPCC trace.

It can be seen that the logarithmic and constant time implementations greedy and const_greedy of the greedy algorithm have the qualitative same behavior. The same is true for the CB and the fast CB strategies.

The greedy strategy, has by far the worst characteristics. Its average WA is still increasing after more than $0.8 \cdot 10^9$ accesses. The lowest write amplification results are achieved by CB and fast CB. They are more than three times smaller than the WA of the greedy strategies, which can significantly increase an SSD's life-time.

The WA of FeGC is in the same order as the amplification of the CB strategies. FeGC looks slightly worse for the TPCC benchmark, but achieves slightly better results for Financial1. FeGC only outperforms the other strategies for the SPC1 trace. CAT, on the other hand, includes a component to improve wear-leveling which seems to have a strong negative impact on write amplification.

Modern SSDs use multiple channels which allow parallel accesses. Nevertheless, complete channels are blocked during GC and the controller always has to select a victim from a specific channel if this channel runs out of blocks, reducing the flexibility of the controller. Fig. 6 shows the WA on a four channel SSD with 2,048 pages per block and 1,024 GByte capacity.

A comparison with Fig. 5a shows the impact of multiple channels on WA for the different strategies. The figures show similar curves where the greedy and the CAT strategy have the highest number of additional write operations. However, for all strategies there is an increase in WA when four channels are used.

We conclude from this first experiments that the WAs of CB are better than (or at least nearly as good as) the one of competing strategies and hereafter focus on the performance of the different approaches.

### 4.4 Performance Results

Fig. 7 shows simulation results for a 512 GByte SSD with different numbers of pages per block. The IOPS are calculated by accumulating the times relevant for the

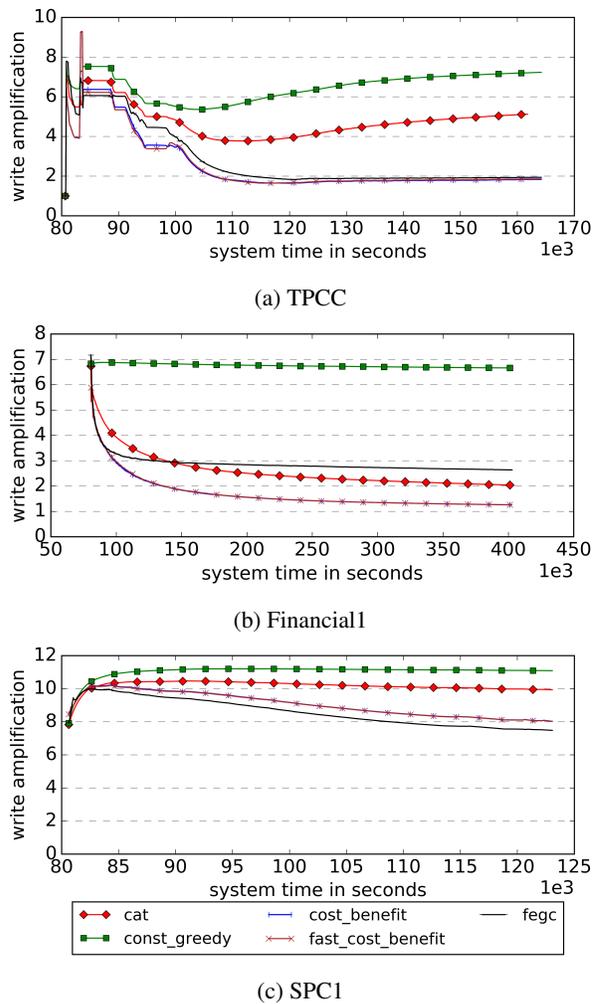

(a) TPCC

(b) Financial1

(c) SPC1

Figure 5: WA for the different GC algorithms and traces

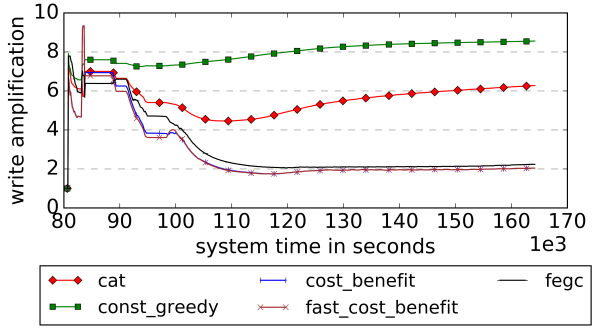

Figure 6: WA results on a four channel SSD for TPCC.

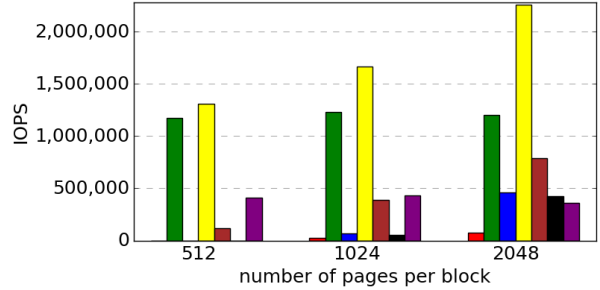

(a) TPCC

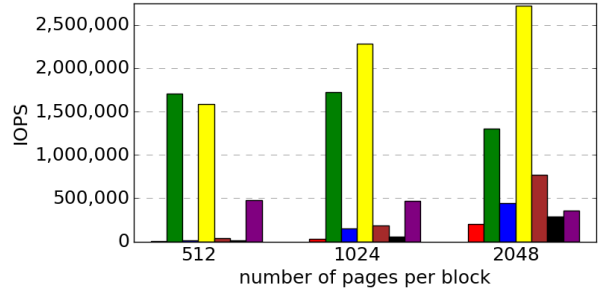

(b) Financial1

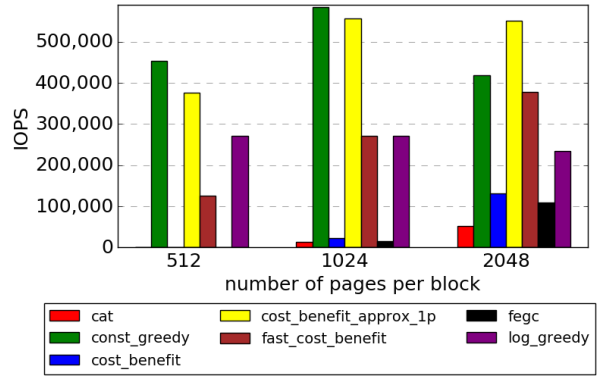

(c) SPC1

Figure 7: Performance results for a capacity of 512 GB and a variable number of pages per block

address translation and GC within the FTL and dividing it by the number of write operations inside the traces.

We assume that all modern SSDs will be able to perform more than 1,000,000 IOPS soon and therefore set a target threshold of 1,800,000 IOPS for the GC strategies alone [20, 22]. For this reason, we terminated all simulations that were unfinished after a runtime of 5 days. This would translate into less than 6,900 IOPS for TPCC, 175 IOPS for Financial1, and 4,400 IOPS for SPC1.

The runtimes of all strategies, besides the constant time greedy implementation, include a logarithmic or linear part, which depends on the number of blocks of an SSD. It is easy to see that the bigger the size of an SSD, the longer the runtime of the GC algorithm. A hidden parameter is the *number of pages per block*. It is possible to reduce the number of blocks of an SSD by increasing the number of pages per block, which then immediately decreases the runtime of the GC strategies.

Fig. 7a shows that for the TPCC trace none of the investigated strategies is able to pass the threshold of 1,800,000 IOPS for a setting with 512 pages per block; even the constant greedy strategy stays below 1,200,000 IOPS and approximative CB[5] stays below 1,400,000 IOPS. Also, only approximative CB is able to use 2,048 pages per block to pass the threshold.

The linear runtime algorithms CB and FeGC perform very badly and have not been able to finish the simulation runs in five days for 512 pages per block, while offering at least a partially acceptable performance for 1,024 and 2,048 pages per block. CAT behaves even worse than CB and FeGC. For TPCC and 2,048 pages per block, the overall simulation runtime for CAT is 38,638 seconds and therefore 6 times longer than the simulation time for the original CB strategy. This cannot be explained by the victim selection strategy alone because the major difference between both implementations is one additional division by the number of erase cycles. It is therefore

---

[5]the suffix _1p in the legend denotes that the number $q$ of cached entries for the approximative cost-benefit is set to 1 % of the number of blocks or 5,608 blocks for a 512 GByte SSD with 512 pages per block

important to include the WA (see Fig. 5) into the overall calculation, as the WA and also the number of GC cycles of CAT is by a factor of 2.4 higher than of CB.

The bad WA also partly explains the performance of constant greedy, as this strategy has 3.2 times more copy operations (in this setting) than approximative CB. For the SPC1 trace constant greedy outperforms approximative CB when the number of pages per blocks is below 2,048. In this case, approximative CB's advantage in the write amplification cannot compensate the faster victim selection of constant time greedy.

Fast CB produces much better results than the original strategy, especially for a small number of pages per block. For 2,048 pages per block, fast CB is even faster than the logarithmic time implementation of the greedy strategy, which has been used here as an example of logarithmic time strategies. A more detailed explanation of the fast cost-benefit runtime will be given in Section 4.6.

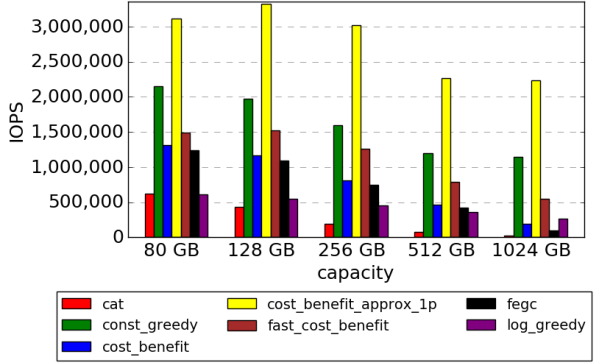

Figure 8: Performance results for the TPCC traces with 2,048 pages per block and variable capacities

The results presented in Fig. 7a indicate that most SSD vendors will apply some approximation of the "pure" strategies presented in this paper.

Fig. 8 shows performance results for the TPCC benchmark, a fixed number of pages per block, and different capacities. The figure helps to understand how the performance of the GC strategies has become more important over time. The constant time implementation for greedy and approximative CB are faster than the threshold of 1,800,000 IOPS for an SSD capacity of 80 and 128 GByte and 2,048 pages per block. Constant time greedy falls below the threshold for a capacity above 256 GByte. Approximative CB is in these settings always faster than the threshold value and achieves up to 3,400,000 IOPS. The performance improvement for approximative CB and 128 GByte SSDs compared to 80 GByte SSDs can be explained by the WA, as the strategy benefits from the lower access densities of the TPCC benchmark above 80 GByte.

Fast CB achieves up to 1,500,000 IOPS, but is unable to achieve the required IOPS performance for any SSD capacity. Nevertheless, it degrades more gracefully than CB or FeGC, while it is always achieving a better performance than the competitors. These results are in line with the theoretical analysis in Subsection 3.1.

An increase of the WA leads to an IOPS decrease when more channels are used (see Fig. 9). A comparison with Fig. 8 shows a performance loss introduced by the additional channels. It has to be considered that our simulator works sequentially and thus it does not exploit the parallelism of the SSD structure. It is therefore possible to parallelize the GC process to four cores and (in the best case) to quadruple IOPS at the cost of more resources. However the comparison of the different strategies shows

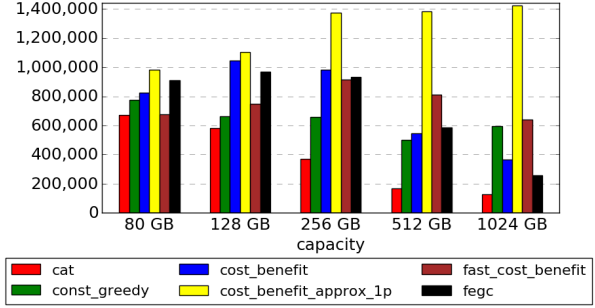

Figure 9: Performance results for the TPCC traces with 2,048 pages per block and four channels.

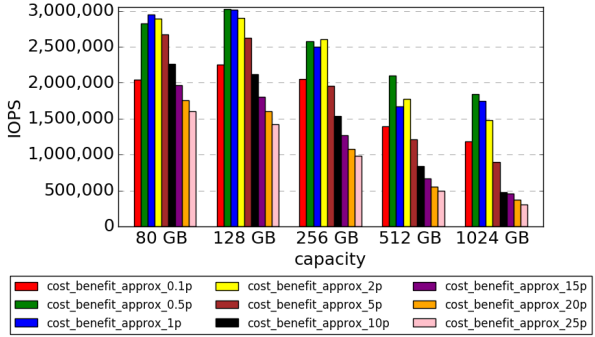

Figure 10: Performance results for approximative CB with 1,024 pages per block.

that the usage of fast CB and approximative CB achieves the best results for large SSDs.

### 4.5 Parameterizing approximative CB

This subsection discusses the influence of the number of candidate blocks $q$ on the performance of the approximative cost-benefit strategy. Fig. 10 compares the performance for the values $q = 0.1\%, 0.5\%, 1\%, 2\%, 5\%, 10\%, 15\%, 20\%, 25\%$. We can derive two opposing trends from these simulations:

1. Increasing $q$ leads to fewer scans for the set of blocks.

2. Increasing $q$ prolongs every scan of the cached blocks as the its average number grows with $q$.

Accordingly, one can observe a performance benefit for $q \in [0.1\%, 2\%]$ in the case of bigger capacities. The results in the previous sections are for $q = 1\%$. This setting is only slightly faster than approximative CB with $q = 0.5\%$ for a capacity of 80 GByte, whereas $q = 0.5\%$ achieves the best performance in all other cases.

The remaining question is whether increasing $q$ also increases WA. Fig. 11 shows that this is the case. The results for approximative CB are becoming significantly worse than the original CB for $q \geq 2\%$ and the setting for $q = 25\%$ already increases WA by a factor of two.

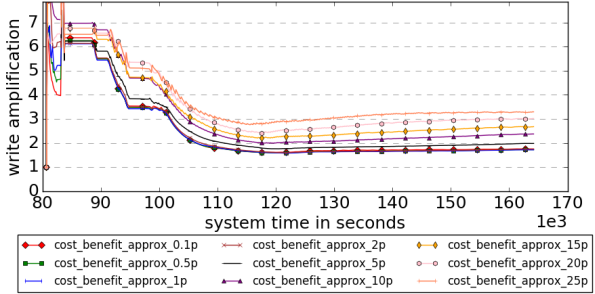

Figure 11: Write amplification for approximative CB with a capacity of 1 TByte and 2,048 pages per block

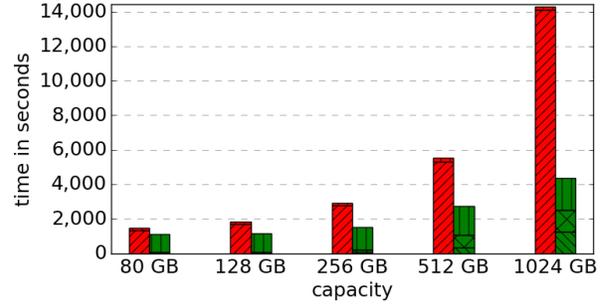

(a) TPCC

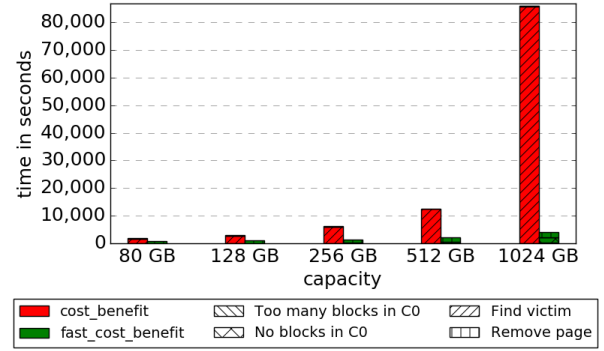

(b) SPC1

Figure 12: Performance analysis for the different traces with 2,048 pages per block and variable capacities.

It is obvious that a larger fraction of cached blocks leads to higher search costs for the victim selection, while reducing the number of times that the cache has to be refreshed. The simulations show that for a 1,024 GByte SSD and $q = 0.1\%$, 99% of the time is spent rebuilding the cache, while for $q \geq 2\%$, between 80% and 94% of the time is spent searching the next victim. This tradeoff again leads to a nice opportunity to parallelize the approximative cost-benefit strategy. For $q = 1\%$, this fraction is 57%, so that there is again the potential to decouple both tasks and to parallelize the strategy.

## 4.6 Evaluation of fast cost-benefit

In the following we analyze the costs of the different operations of CB and fast CB, which are based, in contrast to approximative cost-benefit, on exactly the same cost function. Since TPCC and Financial1 behaved similarly, we only present results for TPCC and SPC1. The measurements illustrate the algorithm's accumulated latencies for managing the classes when the operations are performed sequentially and synchronously.

Fig. 12 shows for each test one bar for CB and fast CB, which is divided into the different functions performed by the strategies. The important functions for CB are finding a victim and removing a page from a block. Fast CB also has to update its classes in the cases that there are no or too many blocks in Class 0.

The figure shows that CB spends most of its time on determining the next victim block, while fast CB which spends most of its time on removing pages. Removing pages in fast CB requires more operations than in the CB strategy because it always has to update Class 1 and potentially move a block from Class 0 to Class 1. For SPC1, fast cost-benefit is up to 13-times faster than cost-benefit, while it is only 2.8-times faster for the TPCC benchmark. The reason is that the size of Class 0 has to be updated less frequently. It can also be noticed that the fraction of the runtimes for changing the size of Class 0 grows with an increasing SSD size.

Figure 12 also reveals an important parallelization potential. Updating Class 1 is independent of the victim selection, so that nearly all costs for removing a page can be asynchronously performed by another core. Also, it is possible to decouple the functions to change the number of blocks in Class 0 so that it is possible to only limit the performance of fast CB based on the critical path length, which would lead to nearly 1,500,000 IOPS (see Fig. 8).

## 4.7 ARM Tests

This section discusses the performance of the strategies on an ARM Cortex-A17 Quad-Core processor with 1.8 GHz implementing the ARM v7 ISA. Since our platform only provides 4 GByte of main memory, we only simulated an 80 GByte SSD with 2,048 pages per block.

Fig. 13 shows results for these settings on an ARM and an Intel processor and for Financial1. The strategies behave similarly on both architectures, but there are also interesting variations. The results show a large performance difference of up to 20-times for the fastest algorithm. This can be only partly explained by the ARM processor's frequency, which is 1.2 times lower than the Intel processor's frequency. Additionally, the Intel Xeon processor provides a 64 bit quad channel DDR4-2133 MHz connection to its main memory while the ARM processor is only equipped with 32 bit dual channel DDR3-1333 MHz RAM.

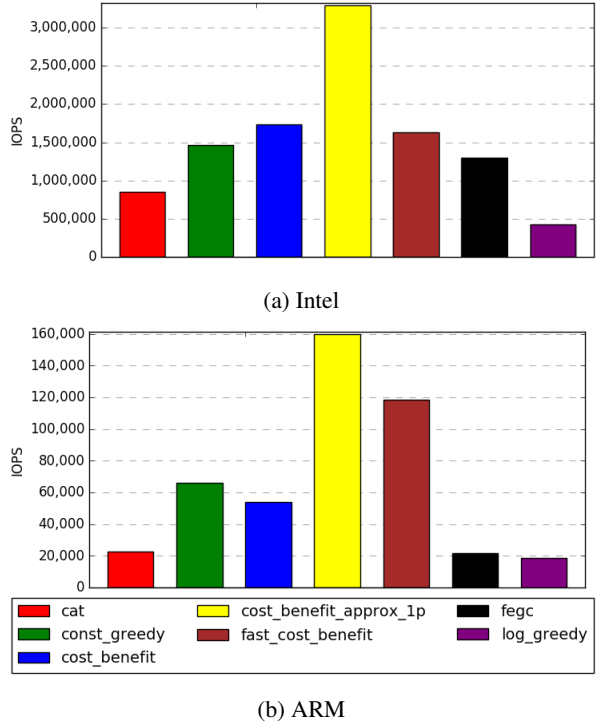

(a) Intel

(b) ARM

Figure 13: Performance results for trace file Financial1 with 2,048 pages per block and a capacity of 80 GB.

The deviations for the different strategies must be explained separately. The difference in the performance of the CB strategy can be explained by the weak floating point performance of the ARM processor. For every victim selection, CB must compute the double precision floating point CB ratio for all blocks. On the other hand, the constant time greedy implementation does not suffer from this limitation because no floating point operations are required. Even though fast CB is influenced by this limited performance, the impact is smaller because it updates fewer elements during victim selection.

FeGC suffers from the ARM architecture, too, but for other reasons. Due to the lower number of channels and the lower clock frequency the memory bandwidth is smaller on the ARM system. FeGC must check the invalidation history of all pages for victim selection and therefore loads all blocks and all pages so that here the memory bandwidth is a limiting factor.

## 5 Conclusion

In this paper we have introduced new GC algorithms that follow the CB approach, but are considerably faster. They can help to significantly improve SSDs' lifetimes by making sophisticated GC strategies fast enough for big SSDs. Fast CB reduces the number of computations by grouping blocks into classes, while the approximative strategy is selecting more than one victim at a time.

Our simulations show that fast CB increases IOPS by a factor of up to 12.1, while preserving the exact same WA. The IOPS achieved by the approximative strategy can be even 24.4 times higher, and it is very surprising that its WA does not noticeably increase for small $q$.

It is important to further improve the performance of our algorithms for even bigger SSDs. We plan to focus on the parallelization potential of the algorithms by partitioning the address space to independently distribute the load to more cores and to optimize the parameters of the strategies. One approach is to use the parallelization potential of multi-channel SSDs by using one independent thread for each channel.

## References

[1] A. Ban, "Wear leveling of static areas in flash memory," May 4 2004, uS Patent 6,732,221. [Online]. Available: https://www.google.com/patents/US6732221

[2] L. Chang, "On efficient wear leveling for large-scale flash-memory storage systems," in *Proceedings of the 2007 ACM Symposium on Applied Computing (SAC), Seoul, Korea*, 2007, pp. 1126–1130.

[3] L. Chang, Y. Liu, and W. Lin, "Stable greedy: Adaptive garbage collection for durable page-mapping multichannel ssds," *ACM Trans. Embedded Comput. Syst.*, vol. 15, no. 1, pp. 13:1–13:25, 2016.

[4] M. Chiang and R. Chang, "Cleaning policies in mobile computers using flash memory," *Journal of Systems and Software*, vol. 48, no. 3, pp. 213–231, 1999.

[5] Y. Deng, "What is the future of disk drives, death or rebirth?" *ACM Computing Surveys (CSUR)*, vol. 43, no. 3, p. 23, 2011.

[6] P. Desnoyers, "Analytic modeling of SSD write performance," in *The 5th Annual International Systems and Storage Conference, SYSTOR '12, Haifa, Israel, June 4-6, 2012*. ACM, 2012, p. 14.

[7] A. Gupta, Y. Kim, and B. Urgaonkar, "DFTL: a flash translation layer employing demand-based selective caching of page-level address mappings," in *Proceedings of the 14th International Conference on Architectural Support for Programming Languages and Operating Systems, ASPLOS 2009, Washington, DC, USA, March 7-11, 2009*, 2009, pp. 229–240.

[8] X.-Y. Hu and R. Haas, "The fundamental limit of flash random write performance: Understanding, analysis and performance modelling," IBM Research Report, 2010/3/31, Tech. Rep., 2010.

[9] S. Im and D. Shin, "Comboftl: Improving performance and lifespan of MLC flash memory using SLC flash buffer," *Journal of Systems Architecture - Embedded Systems Design*, vol. 56, no. 12, pp. 641–653, 2010.

[10] X. Jimenez, D. Novo, and P. Ienne, "Wear unleveling: improving NAND flash lifetime by balancing page endurance," in *Proceedings of the 12th USENIX conference on File and Storage Technologies, FAST 2014, Santa Clara, CA, USA, February 17-20, 2014*, 2014, pp. 47–59.

[11] S. Jung and Y. H. Song, "LINK-GC: a preemptive approach for garbage collection in NAND flash storages," in *Proceedings of the 28th Annual ACM Symposium on Applied Computing, SAC '13, Coimbra, Portugal, March 18-22, 2013*, 2013, pp. 1478–1484.

[12] A. Kawaguchi, S. Nishioka, and H. Motoda, "A flash-memory based file system," in *Proceedings of the 1995 Technical Conference on UNIX and Advanced Computing Systems (USENIX), New Orleans, Louisiana, USA*, 1995, pp. 155–164.

[13] J. Kim, J. M. Kim, S. H. Noh, S. L. Min, and Y. Cho, "A space-efficient flash translation layer for compactflash systems," *IEEE Trans. Consumer Electronics*, vol. 48, no. 2, pp. 366–375, 2002.

[14] O. Kwon, K. Koh, J. Lee, and H. Bahn, "Fegc: An efficient garbage collection scheme for flash memory based storage systems," *Journal of Systems and Software*, vol. 84, no. 9, pp. 1507–1523, 2011.

[15] Y. Li, P. P. C. Lee, and J. C. S. Lui, "Stochastic modeling and optimization of garbage collection algorithms in solid-state drive systems," *Queueing Syst.*, vol. 77, no. 2, pp. 115–148, 2014.

[16] S.-P. Lim, S.-W. Lee, and B. Moon, "Faster ftl for enterprise-class flash memory ssds," in *Proceedings of the 2010 International Workshop on Storage Network Architecture and Parallel I/Os (SNAPI)*, 2010, pp. 3–12.

[17] F. Margaglia, G. Yadgar, E. Yaakobi, Y. Li, A. Schuster, and A. Brinkmann, "The devil is in the details: Implementing flash page reuse with WOM codes," in *14th USENIX Conference on File and Storage Technologies, FAST 2016, Santa Clara, CA, USA, February 22-25, 2016.*, 2016, pp. 95–109.

[18] J. Menon and L. Stockmeyer, "An age-threshold algorithm for garbage collection in log-structured arrays and file systems," in *High Performance Computing Systems and Applications*, 1998, pp. 119–132.

[19] M. Rosenblum and J. K. Ousterhout, "The design and implementation of a log-structured file system," *ACM Trans. Comput. Syst.*, vol. 10, no. 1, pp. 26–52, 1992.

[20] E. Seppanen, M. T. O'Keefe, and D. J. Lilja, "High performance solid state storage under linux," in *IEEE 26th Symposium on Mass Storage Systems and Technologies, MSST 2012, Lake Tahoe, Nevada, USA, May 3-7, 2010*, 2010, pp. 1–12.

[21] R. Stoica and A. Ailamaki, "Improving flash write performance by using update frequency," *PVLDB*, vol. 6, no. 9, pp. 733–744, 2013.

[22] V. Vasudevan, M. Kaminsky, and D. G. Andersen, "Using vector interfaces to deliver millions of IOPS from a networked key-value storage server," in *ACM Symposium on Cloud Computing, SOCC '12, San Jose, CA, USA, October 14-17, 2012*, 2012, pp. 8:1–8:13.

[23] M. Wu and W. Zwaenepoel, "envy: A non-volatile, main memory storage system," in *Proceedings of the Sixth International Conference on Architectural Support for Programming Languages and Operating Systems (ASPLOS), San Jose, California, USA*, 1994, pp. 86–97.

[24] J. Zhang, J. Shu, and Y. Lu, "Parafs: A log-structured file system to exploit the internal parallelism of flash devices," in *2016 USENIX Annual Technical Conference, USENIX ATC 2016, Denver, CO, USA, June 22-24, 2016.*, 2016, pp. 87–100.